\documentclass[11pt]{article}

\usepackage{fullpage}
\usepackage{epic,eepic,epsfig,graphics}
\usepackage{amsmath}
\usepackage{color}
\usepackage{latexsym}
\usepackage{tabularx}
\usepackage{times}
\usepackage{graphicx}
\usepackage{epstopdf}

\graphicspath{{figures/}}

{

\newcommand{\toto}{xxx}

\newenvironment{theorem-repeat}[1]{\begin{trivlist}
\item[\hspace{\labelsep}{\bf\noindent Theorem~\ref{#1} }]}%
{\end{trivlist}}

\newenvironment{lemma-repeat}[1]{\begin{trivlist}
\item[\hspace{\labelsep}{\bf\noindent Lemma~\ref{#1} }]}%
{\end{trivlist}}

\newcounter{linecounter}
\newcommand{\linenumbering}{(\arabic{linecounter})}
\renewcommand{\line}[1]{\refstepcounter{linecounter}
\label{#1}
\linenumbering}
\newcommand{\resetline}{\setcounter{linecounter}{0}}

\title{\bf Tracking COVID-19 by Tracking Infectious Trajectories: An IoT Investigation System\footnote{The authors declare that they have no known competing for financial interests or personal relationships that could have appeared to influence the work reported in this paper.}} 

\title{\bf Tracking COVID-19 by Tracking Infectious Trajectories\footnote{The authors declare that they have no known competing for financial interests or personal relationships that could have appeared to influence the work reported in this paper.}} 

\author{Badreddine Benreguia, Hamouma Moumen\footnote{Correspondence: Dr. Hamouma Moumen. Email: hamouma.moumen@univ-batna2.dz}, and Mohammed Amine Merzoug\\
Computer Science Dept., University of Batna 2\\
53, Road of Constantine, Fesdis\\
Batna 05078, Algeria\\
\{badreddine.benreguia, hamouma.moumen, amine.merzoug\}@univ-batna2.dz}

\date{}

\begin{document}

\maketitle

\begin{abstract}

Nowadays, the coronavirus pandemic has and is still causing large numbers of deaths and infected people. Although governments all over the world have taken severe measurements to slow down the virus spreading (e.g., travel restrictions, suspending all sportive, social, and economic activities, quarantines, social distancing, etc.), a lot of persons have died and a lot more are still in danger. Indeed, a recently conducted study~\cite{ref2} has reported that 79\% of the confirmed infections in China were caused by undocumented patients who had no symptoms. In the same context, in numerous other countries, since coronavirus takes several days before the emergence of symptoms, it has also been reported that the known number of infections is not representative of the real number of infected people (the actual number is expected to be much higher). That is to say, asymptomatic patients are the main factor behind the large quick spreading of coronavirus and are also the major reason that caused governments to lose control over this critical situation. To contribute to remedying this global pandemic, in this paper, we propose an IoT\footnote{IoT: Internet of Things.} investigation system that was specifically designed to spot both undocumented patients and infectious places. The goal is to help the authorities to disinfect high-contamination sites and confine persons even if they have no apparent symptoms. The proposed system also allows determining all persons who had close contact with infected or suspected patients. Consequently, rapid isolation of suspicious cases and more efficient control over any pandemic propagation can be achieved.\\


\end{abstract}


\section{Introduction}
\label{sec:intro}

In December 2019, a novel virus has emerged in Wuhan city. This disease, named coronavirus or COVID-19, has quickly spread throughout China and then to the entire world. As of May 2020, more than 200 countries and territories have been affected and over 4 million people have been diagnosed with the virus. Since no cure or vaccine has been found and since tests cannot be applied on a large scale to millions of persons, governments had and still have no choice but to take severe actions such as border closing, travel canceling, curfews, quarantines, and contact precautions (facemasks, social distancing, and self-isolation). The authorities have also implemented strategies that aim to rapidly detect infections using different cutting-edge medical tools (such as thermal cameras, blood tests, nasal swabs, etc.). On the one hand, these measurements, which for the time being constitute the only possible solution, have succeeded to slow down the contagion spreading, but on the other hand, they have also caused considerable economic damages, especially to countries with brittle economies (suspension of all sorts of activities: social, economic, educational, etc.).        

\begin{figure}[!b]
\begin{center}
\includegraphics[width=5.8in]{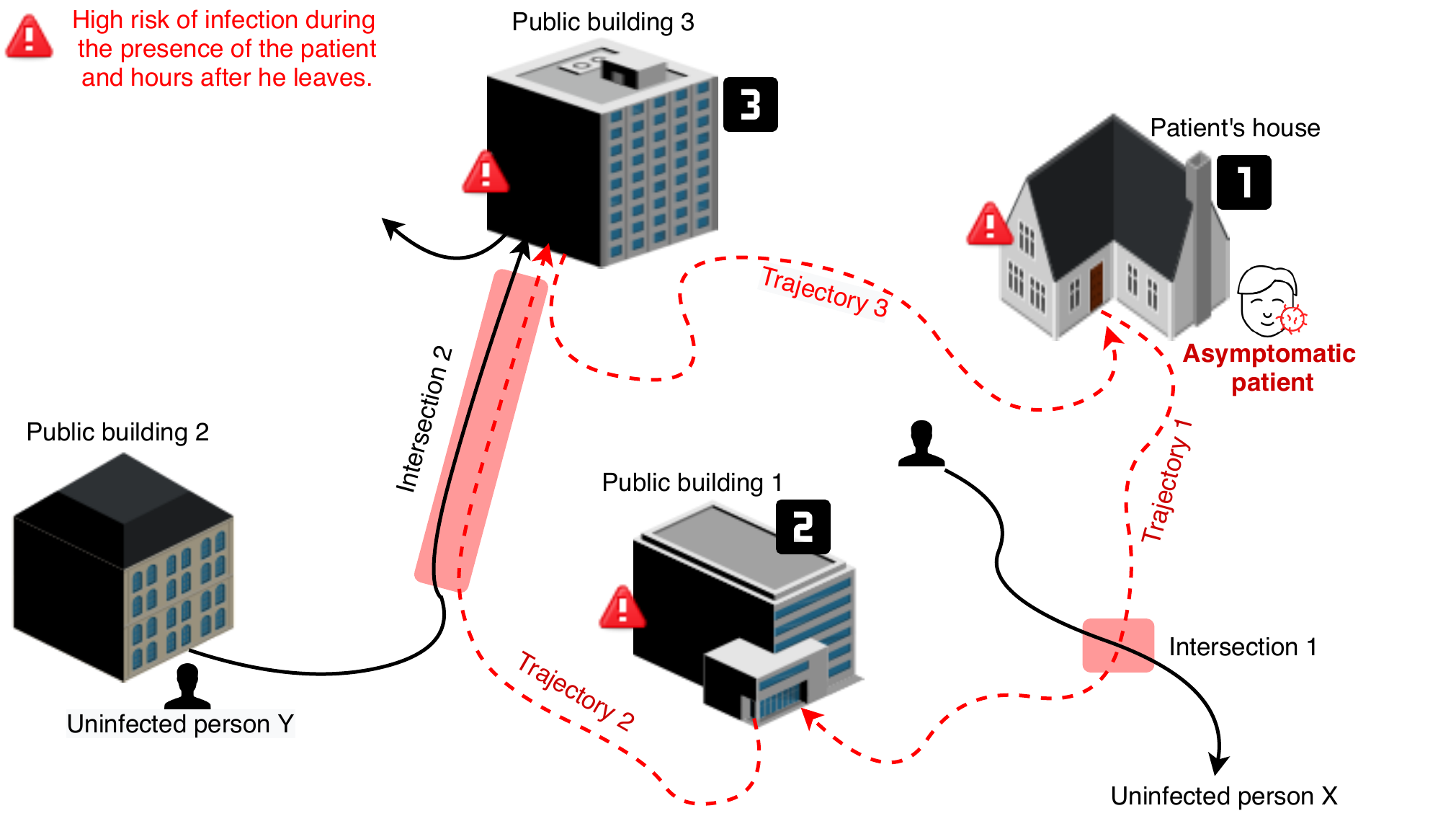}
\caption{Infection tracking: example of historical trajectories taken by a COVID-19 patient and all persons (resp. places) who he might have encountered/infected (resp. visited).}
\end{center}
\label{fig:trajectory-tracking}
\end{figure}

The rapid spreading of coronavirus is due to the continuous person-to-person transmission~\cite{ref1,ref3,ref4}. In addition to this, a recent study has also suggested that a second factor is playing a major causal role in this high virus spreading; namely the \textit{stealth transmission}~\cite{ref2}. The coronavirus can take 14 days before the appearance of symptoms. During this incubation period, asymptomatic patients, called \textit{undocumented patients}, can infect large communities of people. In turn, these newly infected persons, who will remain unaware of their illness (until they eventually develop the symptoms), can also infect larger communities, thus, leading to an uncontrollable domino effect~\cite{ref2,ref5, ref6}. Accordingly, to confine and effectively eliminate the coronavirus, it is crucial and mandatory to possess an efficient investigation system that allows determining (1)~highly infectious places and (2)~all the persons who were in contact with patients who have recently tested positive. Indeed, persons who are known to be in direct relationship with a patient (such as family members, friends, and coworkers) can be easily determined and tested. However, numerous other persons could have also been in contact with this infected individual. All these persons, who cannot be easily determined, can contribute to the widespread of the virus. To remedy this issue, in this paper, we propose an IoT investigation system that can determine the exact trajectories of infected persons (exact coordinates labeled with time). Hence, as shown in Figure~\ref{fig:trajectory-tracking}, the proposed solution allows determining areas of high contamination and also provides a quick detection mechanism of undocumented patients who are known to be the main and most important factor in the rapid widespread of coronavirus. These persons (resp. places) can be tested/confined even if they had no blatant symptoms (resp. closed and properly disinfected).

The investigation system presented in this paper is a proposition to governments and lawmakers. So, it needs to be approved and adopted only in the case of highly dangerous pandemics and public health emergencies. In summary, to properly operate, the proposed system must be continuously fed with the coordinates of persons who are in public crowded places (mainly using IoT devices that can identify persons and report their locations). Although this proposed solution can open large debates from the standpoint of privacy and human rights, during extreme critical situations in which the entire human race might be at stake, the urgency of saving lives becomes of higher priority. In such circumstances, the proposed technique can be quickly applied as a last resort by higher authorities (governments, WHO\footnote{WHO: World Health Organization.}, etc.) while giving guarantees about (1)~protecting the privacy of people, and (2)~disabling this system once the outbreak is contained. 

The remainder of this paper is organized into five sections as follows. Section~\ref{sec:proposedSolution} describes and details the proposed system. Section~\ref{sec:systemImplemention} presents a small-scale implementation example of this proposal. Section~\ref{sec:mathematical-modeling} addresses the benefits that the proposed investigation system can bring to the state-of-the-art mathematical disease spreading/prediction models. Section~\ref{sec:advantages-disadvantages} discusses the main advantages and shortcomings of the proposed solution. Finally, Section~\ref{sec:conclusion} concludes the paper and provides some recommendations.

\section{Proposed solution}
\label{sec:proposedSolution}

In the proposed system, a big-data architecture is considered to archive the continuously collected trajectories of persons. This archiving structure, which is inspired by the big data model proposed in~\cite{haroun2017big}, must (as previously mentioned) be fed by IoT devices that can (1)~determine coordinates of persons during their outdoor activities and (2)~send the collected data to the system. We point out that persons staying at home or in their vehicles are assumed to be isolated (i.e., they cannot infect other persons nor they can be infected). Consequently, they do not need to activate the coordinates collection process. On the contrary, to ensure their safety by ensuring the proper operation of the system, it is necessary for any person leaving his/her house to activate coordinates collection and save his/her tracks.

\begin{figure}[!t]
\begin{center}
\includegraphics[width=5.2in]{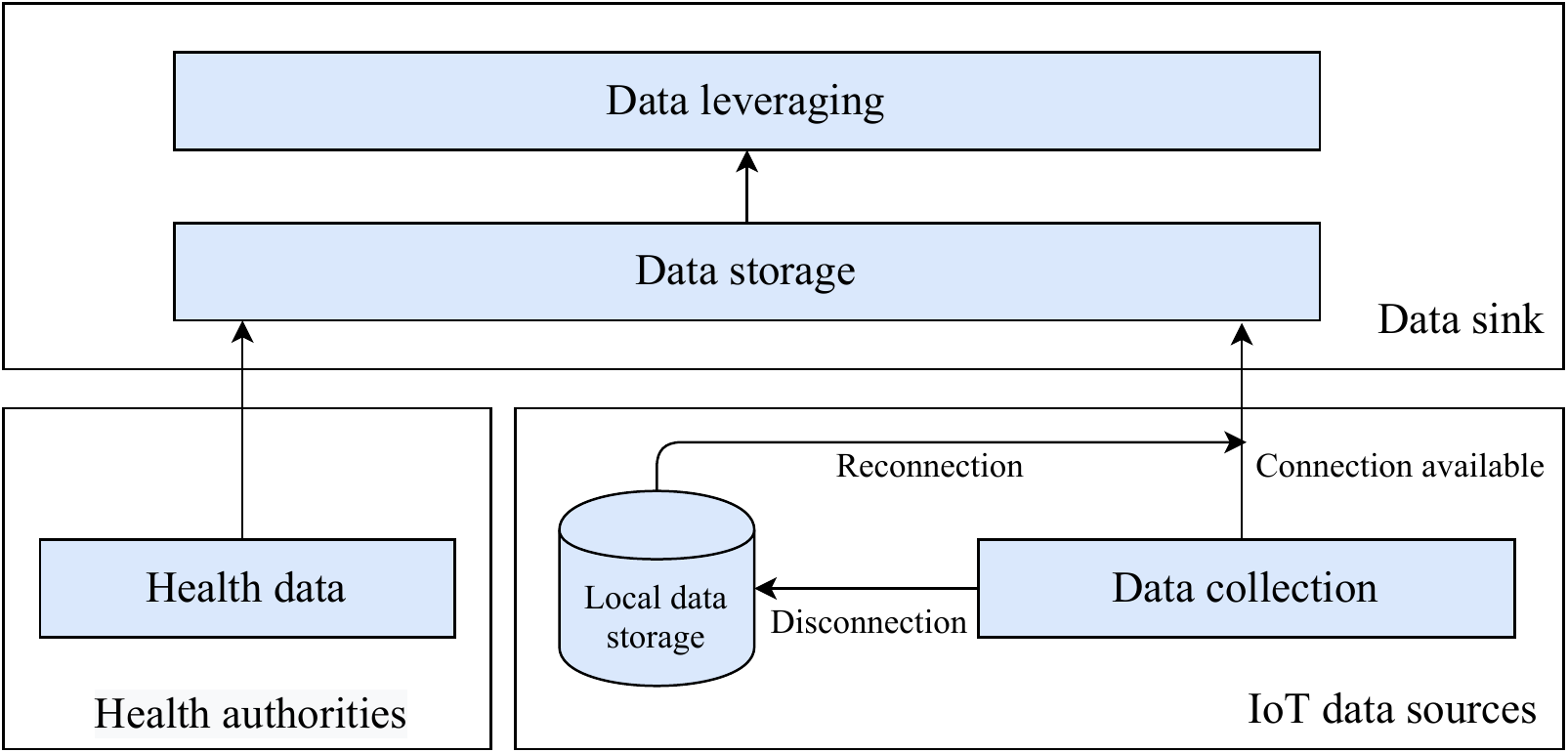}
\caption{Overall system architecture.}
\end{center}
\label{fig:bigDataArchitecture}
\end{figure}

As Figure~\ref{fig:bigDataArchitecture} depicts, the proposed architecture has three main basic layers; namely, data collection, data storage, and data leveraging. In the remainder of this section, we describe each of these layers and provide their specific complementary tasks.

\subsection{Data collection}
\label{sec:data-collection}

As shown in Figure~\ref{fig:bigDataArchitecture}, the data gathering process is divided into two parallel independent tasks. Whereas the first one is related to collecting the necessary health data, the second is responsible for collecting the geolocalization coordinates of persons.

\subsubsection{Health data}
\label{sec:collect-of-health-data}

Reporting the coronavirus deaths and all the newly confirmed and recovered cases is a basic stepping stone in the overall investigation process. For instance, once a new case has been discovered, the latter must be immediately reported to the automatic investigation system to allow it to find all the individuals who could have been infected by this person. In the same context, given the fact that patients are the main factor in the system, once some of them have recovered from the virus, their statuses must be immediately changed from active to recovered, and so forth. 

On the contrary to the second task (i.e., geolocalization data gathering) in which data is automatically captured and reported by the employed IoT devices, in this first task, data must be manually entered by health authorities. For example, hospitals can be responsible for providing the newly infected cases, and all the other indispensable updates. 

\subsubsection{Geolocalization data}
\label{sec:collect-of-geolocalization-data}

Different approaches can be considered to collect the coordinates of persons who left their houses. For example, (1)~relying on telecom and IT infrastructures, (2)~equipping public buildings with appropriate devices, or (3)~utilizing dedicated tracking apparatuses. 
 
\begin{itemize}

\item \textit{Telecom/high-tech companies}: using telecommunication technologies, smartphones can be easily instructed (programmed) to periodically report their actual location to a dedicated storage system (phone-number, GPS-coordinates, time).

\item \textit{Public buildings}: by installing face recognition cameras on the entries/exits of public buildings, the identity of visitors can be easily determined and then sent to the designated infrastructure (person-id, building-id, entry-time, exit-time). This way, based on the information provided by the different public buildings and facilities, the tracks of infected people can be easily determined. The following lines provide an example of places that were visited by an infected person \texttt{P1}:\\

\texttt{(P1, building, (entry, T1), (exit, T2)),\\ 
(P1, shopping mall, (entry, T3), (exit, T4)),\\
(.., .., (.., ..), (.., ..)),\\
(P1, airport A, (entry, Tm), (exit to plane, Tn)),\\
(P1, airport B, (entry from plane, Tx), (exit, Ty)),\\
(.., .., (.., ..), (.., ..).\\
} 

The main drawback of this second data collection approach is that it cannot be easily implemented in most countries due to the lack of appropriate underlying identification/recognition systems. However, this solution can be utilized in the case where authorities fail to convince people of using their cellphones as tracking devices.

\item \textit{Specific solutions and electronic devices}: the two aforementioned approaches can be adopted independently or in combination. Actually, during dangerous outbreaks where rapid tracking of infected people becomes an urgency, in addition to personal smartphones and public indoor (entry/exit) cameras, governments can consider using other tracking techniques like for instance electronic bracelets, public outdoor security cameras, drones, or even satellites. Furthermore, to consolidate the previous two approaches, solutions such as Bluetooth or NFC\footnote{NFC: Near-Field Communication.} can also be considered~\cite{ref1_tracking,ref2_tracking,ref3_tracking,ref4_tracking,ref5_tracking}. In the same context, in the event of technical issues related to GPS, governments can opt for alternative efficient geo-localization techniques~\cite{ref2_tracking,ref4_tracking}. 

\end{itemize}

As their name implies, the IoT devices (responsible for collecting person tracks) report the required data using the Internet (via wired or mobile wireless networks: 4/5G, ...). So, as shown in Figure~\ref{fig:bigDataArchitecture}, to avoid losing important information in the case of Internet disconnection, each device must store locally the collected data. Once reconnected, these IoT devices can then send the gathered data to the system. Finally, it is worth mentioning that to ensure the success of the whole virus tracking process, IT companies, public institutions, and third-parties must contribute to collecting and reporting the required data.

\subsection{Data storage}
\label{subsec:data-storage}

Big data, which refers to enormous datasets, has five essential characteristics known as the 5~Vs: volume (data size), velocity (data generation frequency), variety (data diversity), veracity (data trustworthiness and quality), and finally value (information or knowledge extracted from data). To address the issues related to storing and processing huge data volumes, extensive research has been done and numerous efficient solutions have been proposed (data ingestion, online stream processing, batch offline processing, distributed file systems, clustering, etc.). These techniques provide both efficient distributed storage and fast processing. Currently, research is more focused on big data leveraging (i.e., the last V which consists of turning data into something valuable using data analytics, machine learning, and other techniques)~\cite{haroun2017big,Johanson2014}. This point, which represents the main contribution of our work, will be detailed in the next subsection.

As stated in~\cite{haroun2017big}, an ideal big data storage/processing infrastructure must adhere to the four following major requirements: (1)~time and space efficiency, (2)~scalability, (3)~robustness against failures and damages, and (4)~data security/privacy. For instance, regarding this last point, since the collected data in our context is personal and highly sensitive, strict laws and regulations must be imposed to protect it. The established privacy model must determine who can access data (governments, higher health authorities, WHO organization, etc.), under what constraints, for how long data can be kept, and to whom this data can be distributed, etc. Moreover, in addition to data privacy, all conventional data security mechanisms must also be considered (confidentiality, integrity, and availability). 

To ensure the proper operation of the system, data that is continuously coming from the various considered IoT sources (such as sensors, smartphones, cameras, bracelets, etc.) must provide three main information: \textit{person-id}, \textit{coordinates}, and \textit{time} (Table~\ref{tab:database-example}). The received trajectories of persons are recorded as an immutable discrete set of points (coordinates) labeled with time. For example, according to Table~\ref{tab:database-example}, the recorded traces of person \texttt{P1} along with their corresponding time instants are: (\texttt{c1}, \texttt{t1}), (\texttt{c4}, \texttt{t3}), (\texttt{c5}, \texttt{t4}), (\texttt{c2}, \texttt{t5}), and (\texttt{c3}, \texttt{t6}). Note that in Table~\ref{tab:database-example}, for any given person, time is a monotone function that must keep growing (it cannot go back). Note also that more than one person can exist in the same coordinates at the same time. For example, \texttt{P1} with \texttt{P3} at (\texttt{c2}, \texttt{t5}) and \texttt{P1} with \texttt{P5} at (\texttt{c3}, \texttt{t6}).

\begin{table}
\centering
\caption{Example of collected trajectories.}

\begin{tabularx}{2.7in}{c|c|c} \hline
\texttt{Person-ID} & \texttt{Coordinates}	  & \texttt{Time}\\ 
\hline
\texttt{P1}	 & \texttt{c1} & \texttt{t1} \\ \hline
\texttt{P2}	 & \texttt{c2} & \texttt{t1} \\ \hline	
\texttt{P3}	 & \texttt{c3} & \texttt{t2} \\ \hline
\texttt{P1}	 & \texttt{c4} & \texttt{t3} \\ \hline
\texttt{P1}	 & \texttt{c5} & \texttt{t4} \\ \hline	
\texttt{P1}	 & \texttt{c2} & \texttt{t5} \\ \hline
\texttt{P3}	 & \texttt{c2} & \texttt{t5} \\ \hline
\texttt{P4}	 & \texttt{c6} & \texttt{t5} \\ \hline	
\texttt{P5}  & \texttt{c3} & \texttt{t6} \\ \hline
\texttt{P1}  & \texttt{c3} & \texttt{t6} \\ \hline
\texttt{...} & \texttt{...} & \texttt{...} \\ \hline
\end{tabularx}
\label{tab:database-example}
\end{table}

\begin{figure}[!b]
\begin{center}
\includegraphics[width=3in]{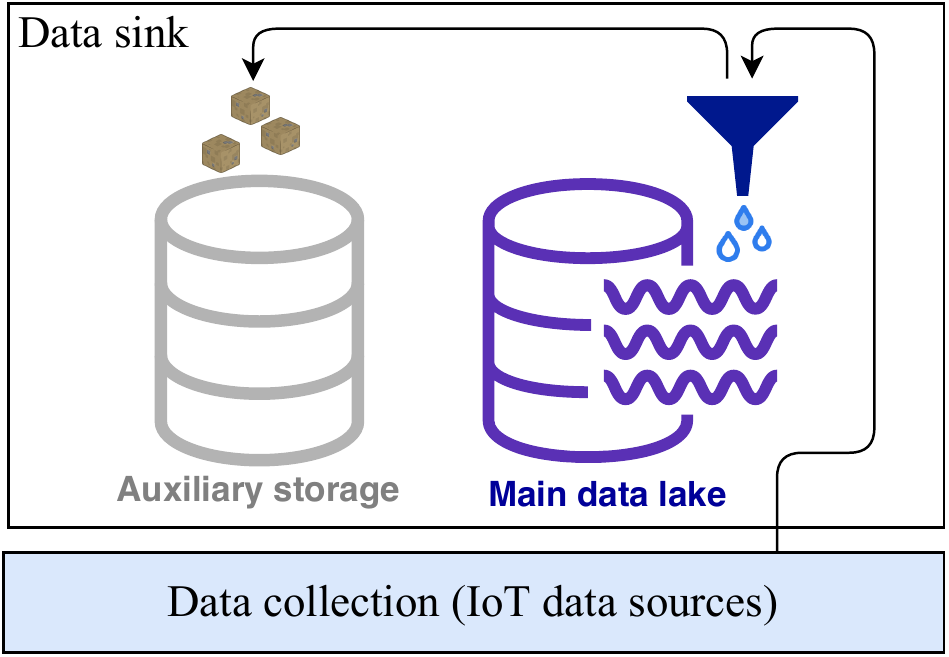}
\caption{Data filtering (preprocessing).}
\end{center}
\label{fig:dataFiltering}
\end{figure}

In the considered context, the \textit{frequency} according to which data must be collected (i.e., the second V) has a direct significant impact on both the first and last Vs (i.e., \textit{data size} and \textit{data value}). High-quality trajectories can be achieved by increasing the frequency of data collection. This, however, can considerably increase the volume of data. In contrast, low collection frequencies reduce the data size but yield low trajectory quality. So, a tradeoff between the quality of trajectories and the size of data must be defined. 

The main goal of the proposed system is to track the main sources of infection, whether they would be humans or places. But, in reality, the quality of the collected data can have a deep impact on the system's performance; it can contribute to its efficiency or deficiency. To prevent useless noisy data from harming or affecting the decisions of the investigation process, a filtering mechanism must be added (Figure~3). If it appears that the filtered data is useless, it can then be safely deleted. For example, data collected from highways and main roads is not of value. In this scenario, users are inside their respective cars and the risk of infection is nil (similarly to houses, people inside the same vehicle are considered as isolated). Concretely speaking, the responsibility of data filtering can be left to authorities, which can decide which geographic zones or areas are of interest.

\subsection{Data leveraging}
\label{subsec:data-leveraging}

In the proposed system, only two operations can be performed on the collected geolocalization data: \textit{storage} and \textit{leveraging} (Figure~\ref{fig:bigDataArchitecture}). This section presents our three proposed algorithms, which exploit the gathered data to (1) find and classify suspected cases (i.e., persons who met with confirmed patients or other suspected cases), (2)~determine black areas (zones with high contamination probability), and (3)~find all persons who visited black areas (also considered as probably infected persons). For the convenience of the presentation, Table~2 compiles all the variables utilized by the different proposed algorithms presented respectively in the following three subsections.

\begin{table}[!b]
\begin{center}
\footnotesize
\begin{tabular}{|l|l|l|}
  \hline
   & {\bf Variable} & {\bf Definition}  \\
  \hline
\textbf{Common} &$(P_i,c_i,t_i)$ & A recorded tuple: person identity, coordinates, and the corresponding collection time.\\     
  &$Trajectory$ &  Set of all trajectories. \\ 
  &$Patient$ & Set of all confirmed patients. A patient is defined as $(P_i,t_{i})$.\\
  &$aux$ & A boolean that indicates whether an end-user is a suspected case.\\
  &$S$ & Server side.\\
  &$C$ & Client side.\\

  \hline
  
 \textbf{Algorithm of Figure~\ref{algo:SuspectsFinding}} &$SUSPECTED_{t}$ & Set of all suspected persons at time $t$.\\
  &$suspected_{t,d}$ & Set of suspected persons at distance $d$ and time $t$.\\
  &$d$ & An integer that indicates how close/far is a person to a confirmed patient.\\
  &$k$ & An index to iterate through all distances.\\
  &$CD$ & Current date.\\  
  &$IP$ & Incubation period.\\
  
  \hline  
  
  \textbf{Algorithm of Figure~\ref{algo:black-areas-determination}} & $A$ & Set of all areas $(a_i, \ c_i)$. \\
     
  &$BA$ & Set of all black areas $(ba_i, \ c_i)$. \\
  
  &$count_k$ & Number of patients who have been to area $k$. \\
  
  &$\alpha$ & A threshold to decide whether a location is a black area. \\
  
  \hline  
  
  \textbf{Algorithm of Figure~\ref{algo:persons-in-black-areas}}&$SUSPECT_t$ & Set of all suspected persons who have frequented black areas at time $t$. \\
  &$BA$ & Set of all black areas $(ba_i, \ c_i)$. \\    
  &$suspect_{t,k}$ & Set of suspected persons who have frequented black area $k$ at time $t$. \\
  &$s$ & An index to explore all black areas.\\
    
  \hline
\end{tabular}
\label{tab:variables}
\caption{Variables used in the proposed algorithms.}
\end{center}
\end{table}

\begin{figure}[!t]
\begin{center}
\includegraphics[width=6.5in]{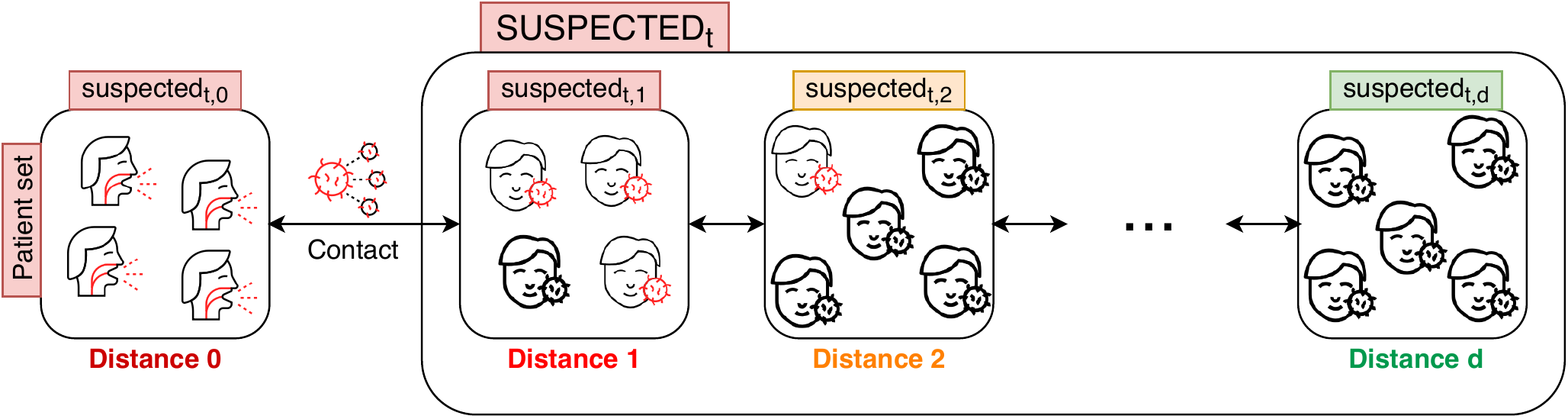}
\caption{Set of all suspected persons at time t.}
\end{center}
\label{fig:SUSPECTED}
\end{figure}

\begin{figure*}[!t]
\centering{ \fbox{
\begin{minipage}[t]{150mm}
\footnotesize
\resetline
\begin{tabbing}

To find and categorize suspected cases into disjoint subsets, the investigation system  $S$ performs the following steps: \\
{\bf Input}:~\= $Trajectory$; $Patient$; \\
{\bf Init}:~\= $d \leftarrow 0$; $t \leftarrow CD$; $SUSPECTED_t\leftarrow \emptyset$; $suspected_{t,d}\leftarrow Patient$; $k \leftarrow 1$;\\
{\bf Begin}\\
\> ------------------------------------------------------------------------------------------------------------- \\

\> ~{\bf whil}\={\bf e } ($suspected_{t,d} \neq \emptyset$) {\bf do}\\
\>\> --------------------------------------------------------------------------------------------------------------- \\
\line{01} \>\>  $d \leftarrow d+1$; \\
\line{07} \>\>  $suspected_{t,d} \leftarrow \emptyset$; \\
\line{07} \>\> {\bf for}\={\bf each} $(person_i, *) \in suspected_{t, d-1}$ {\bf do}\\
\line{07} \>\>\> {\bf if} ($\exists (Person_i,c_i, t_i)\in
Trajectory$ ) {\bf and} ($\exists (Person_j,c_i, t_i)\in
Trajectory$ ) with ($(Person_i, t_i)\notin suspected_{t,d-1}$)\\
~~~~~~~~~~~~~~~~~~~~~~~~{\bf and} $((CD-t_i)\leq IP))$
 {\bf then} $suspected_{t,d}  \leftarrow suspected_{t,d}  \cup \{(Person_j, t_i)\} $;\\
\line{07} \>\> {\bf endfor}\\
\line{07} \>\> $SUSPECTED\leftarrow SUSPECTED \cup  suspected_{t,d}$;\\
\>\> --------------------------------------------------------------------------------------------------------------- \\
\> ~{\bf endwhile}\\

\line{11} \> {\bf upon}\=  ~~  reception of {\sc Query}$(person_i)$  from $c$ {\bf do} \\

\>\> ~{\bf whil}\={\bf e } ($k\leq d-1$) {\bf and}$(Person_i, *)\notin suspected_{t, k-1}$ {\bf do}\\
\>\> --------------------------------------------------------------------------------------------------------------- \\
\line{01} \>\>  {\bf if} $\exists(Person_i, *)\in
suspected_{t,k}${\bf then}~  $aux \leftarrow true$; $s \leftarrow
k$
{\bf else} $aux \leftarrow false$  {\bf endif}\\
\line{01} \>\>  $k  \leftarrow k+1$;\\

\>\> --------------------------------------------------------------------------------------------------------------- \\
\> \>~{\bf endwhile}\\
\line{01} \> \>{\sl send} {\sc Infection}$(aux, s)$ To $c$; \\
{\bf End.}\\

\end{tabbing}
\normalsize
\end{minipage}
} 
\caption{Main investigation algorithm - finding and categorizing suspected cases into disjoint subsets.} 
\label{algo:SuspectsFinding}}
\end{figure*}

\subsubsection{Suspected cases determination}
\label{subsubsec:suspects-determination} 

Based on the stored data (both recorded trajectories of people and health information), the main goal is to deduce all possible infections. As previously mentioned, the proposed investigation system does not only find the suspected cases but also categorizes them into several disjoint subsets denoted $suspected_{t, d}$ (Figure~4 and Table~2). The variable $d$, which stands for distance or degree, indicates how close/far is a person to a confirmed patient. The initial class of persons $suspected_{t, 0}$ represents the set of all patients (confirmed cases). If set $suspected_{t, 1}$ is defined ($suspected_{t, 1} \neq \phi$) then this means that it contains all individuals who had direct contact with at least a confirmed patient $P_c$ from $suspected_{t, 0}$ ($ \forall \ P_c \in suspected_{t, 0}$). As regards the elements (persons) of $suspected_{t, 2}$, they had no direct intersection with a positive case from $suspected_{t, 0}$ (set of confirmed cases), but they have certainly met at least an element from $suspected_{t, 1}$. In general, for $i >= 2$, each element of $suspected_{t, i}$ has certainly met at least an element from $suspected_{t, (i-1)}$, but none from $suspected_{t, (i-2)}$.

To properly operate, the proposed approach must be given the initial set of patients $suspected_{t, 0}$ (suspects at distance $0$). As shown by the algorithm of Figure~\ref{algo:SuspectsFinding}, at each iteration, based on the precedently entered/calculated set $suspected_{t, (i-1)}$, the new set of suspected persons $suspected_{t, i}$ is determined. Note that a person is considered to be suspected if he has met a patient or another suspected person during the incubation period (line 4: $(CD-t_i)\leq IP$). The algorithm stops when the previously entered/calculated set $suspected_{t, (i-1)}$ is empty (which means that the next set of suspected persons $suspected_{t, i}$ cannot be calculated). Thus, after its execution, the proposed algorithm gives a classification (partition) of all individuals stored in the system: $suspected_{t, 0}$, $suspected_{t, 1}$, ..., $suspected_{t, d}$ (\texttt{SUSPECTED}$_t$ set). The remaining unclassified persons are considered uninfected. They have not met with confirmed patients from $suspected_{t, 0}$ nor with suspected persons from $suspected_{t, i}$ (with $i > 0$).

\begin{figure*}[!t]
\centering{ \fbox{
\begin{minipage}[t]{150mm}
\footnotesize
\resetline
\begin{tabbing}

A client $C$ executes the following steps: \\

{\bf Beg}\={\bf in}\\

\> ------------------------------------------------------------------------------------------------------------- \\
\line{01} \> {\sl send} {\sc Query}$(person_i)$ To $S$; \\

\line{01} \> {\bf wait until} (received  {\sc Infection}$(aux,s)$) from $S$ \\

\>-------------------------------------------------------------------------------------------------------------\\
{\bf End.}\\
\end{tabbing}
\normalsize
\end{minipage}
} 
\caption{Algorithm executed by clients (e.g., mobile applications).} 
\label{algo:clients}}
\end{figure*}


When a user (client) sends a query to the investigation system (Algorithm of Figure~\ref{algo:clients}), he will receive to which class he belongs (i.e., $suspected_{t, 0}$, $suspected_{t, 1}$, $suspected_{t, 2}$, ... or $negative$ category). Using the distance $d$ (rather than the identity of persons) allows informing (reassuring or alerting) users about their current statuses without exposing the privacy of other users of the system. The distance $d$ can be seen as a warning, the more a user's distance is close to zero, the higher the risk of infection will be, and vice versa (Figure~4). Finally, we point out that in the proposed system, geolocalization data is periodically collected every $\Delta t$ period. To avoid any synchronization issue that might occur during this process, time differences that are less or equal to $\Delta t$ are ignored. For example, if a person $(P_i, \ c_i, \ t_i)$ has the same coordinates as another person $(P_j, \ c_j, \ t_j)$ and the difference between the collection instants ($t_i$ and $t_j$) of these coordinates ($c_i$ and $c_j$) is less or equal to $\Delta t$ then $ti$ and $tj$ will be seen as equal ($t_i = t_j$).

\begin{figure*}[!t]
\centering{ \fbox{
\begin{minipage}[t]{150mm}
\footnotesize
\resetline
\begin{tabbing}

To find black zones, the investigation system  $S$ performs the following steps: \\
{\bf Input}:~ $Trajectory$; $A$;  $Patient$;   \\
{\bf Init}:~\=  $BA \leftarrow \emptyset$; $s \leftarrow 1$;\\
{\bf Begin}\\
\> ------------------------------------------------------------------------------------------------------------- \\

\line{07} \> {\bf for}\={\bf each} $(a_k,  c_k) \in A$ {\bf do}\\
\line{07} \>\> $count_k  \leftarrow 0$;\\
\line{07} \>\> {\bf if} (($\exists (Person_i, *)\in Patient$ {\bf
and} ($\exists (Person_j,c_k, t_i)\in Trajectory$ ) )
 {\bf then}
 $count_k  \leftarrow count_k  +1 $\\
 \line{01} \>\> {\bf if} ($count_k \geq \alpha$) {\bf then} $BA  \leftarrow BA \cup \{(a_k, c_k)\} $; \\
\line{07} \> {\bf endfor}\\

\> --------------------------------------------------------------------------------------------------------------- \\

{\bf End.}\\

\end{tabbing}
\normalsize
\end{minipage}
} 
\caption{Black areas determination algorithm.}
\label{algo:black-areas-determination}
}
\end{figure*}

\begin{figure}[!t]
\begin{center}
\includegraphics[width=5in]{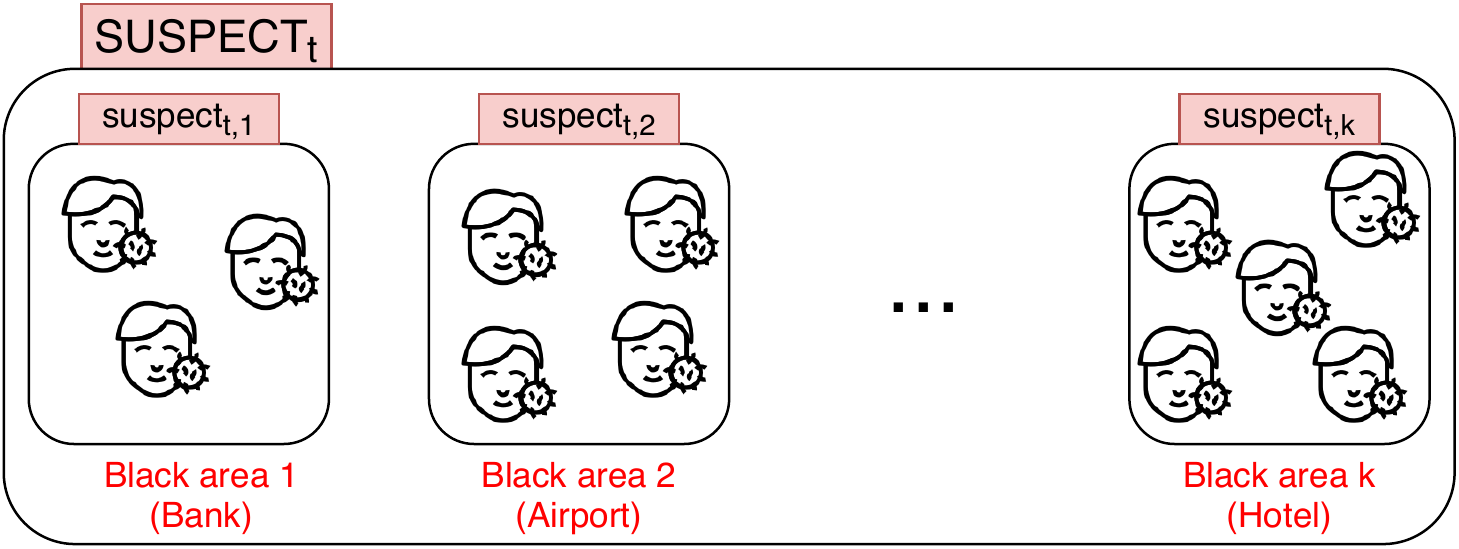}
\caption{Set of all suspected persons who have frequented black areas at time t.}
\end{center}
\label{fig:SUSPECT}
\end{figure}

\begin{figure*}[!t]
\centering{ \fbox{
\begin{minipage}[t]{150mm}
\footnotesize
\resetline
\begin{tabbing}

To find all persons who visited black areas, the investigation system $S$ performs the following steps: \\
{\bf Input}:~ $Trajectory$; $BA$;  \\
{\bf Init}:~\= $t \leftarrow CD$; $SUSPECT_t \leftarrow \emptyset$; $s \leftarrow 1$;\\
{\bf Begin}\\
\> ------------------------------------------------------------------------------------------------------------- \\

\line{07} \> {\bf for}\={\bf each} $(ba_k, c_k) \in BA$ {\bf do}\\
\line{07} \>\> $suspect_{t,k}  \leftarrow \emptyset$;\\
\line{07} \>\> {\bf if} ($\exists (Person_i,c_i, t_i)\in
Trajectory$ ) with$(c_k=c_i)$
 {\bf then}
 $suspect_{t,k}  \leftarrow suspect_{t,k}  \cup \{(Person_i, t_i)\} $\\
 \line{01} \>\>  $SUSPECT_t \leftarrow SUSPECT_t \cup suspect_{t,k}$; \\
\line{07} \> {\bf endfor}\\

\>\> --------------------------------------------------------------------------------------------------------------- \\

\line{11} \> {\bf upon}\=  ~  reception of {\sc Query}$(person_i)$  from $c$ {\bf do} \\

\> ~{\bf whil}\={\bf e } ($s\leq |BA|$) {\bf do}\\
\>\> --------------------------------------------------------------------------------------------------------------- \\
\line{01} \>\>  {\bf if} $\exists(Person_i, *)\in
suspect_{t,s}${\bf then} $aux \leftarrow true$; $k \leftarrow s$
{\bf else} $aux \leftarrow false$  {\bf endif}\\
\line{01} \>\>  $s  \leftarrow s+1$;\\
\>\> --------------------------------------------------------------------------------------------------------------- \\
\> {\bf endwhile}\\
\line{01} \> {\sl send} {\sc Infection}$(aux, ba_k)$ To $c$; \\
{\bf End.}\\

\end{tabbing}
\normalsize
\end{minipage}
} 
\caption{Algorithm for finding persons who have visited black areas.}
\label{algo:persons-in-black-areas}
}
\end{figure*}

\subsubsection{Black areas determination}
\label{subsubsec:black-areas-determination}

Another major problem that must be tackled is the possibility of virus transmission without direct human contact. This can happen, for instance, when an undocumented patient visits a public place and leaves the virus there (he touches objects found in that area, he sneezes, coughs, etc.). In such a scenario, as long as the virus stays alive, the probability of contamination remains very high. The algorithm of Figure~\ref{algo:black-areas-determination} demonstrates the process followed to find all black areas. The key idea consists of determining if numerous patients have visited the same location. If so, then this zone is likely an infectious area of high viral transmission. More concretely, for each public location $a_k \in A$ (where $A$ is the predefined set of all public areas), the number of confirmed patients who have visited this area is determined using the $count_k$ variable. If this number overtakes the predefined threshold $\alpha$, the corresponding location $a_k$ is then considered to be a black area.  

\subsubsection{Suspects from black areas}
\label{subsubsec:suspects-from-black-areas} 

As depicted in Figure~8 and Algorithm of Figure~\ref{algo:persons-in-black-areas}, once black areas have been determined, all persons who have been in these infectious zones can be straightforwardly found. More specifically, based on the set of captured trajectories, the proposed algorithm finds each person $P_i$ who has visited at least one black area $ba_k$. Each of these found suspected people will be added to the corresponding $suspect_{t, k}$ set (list of all suspected persons who have frequented black area $k$ at time $t$). In other terms, after its execution, the algorithm depicted in Figure~\ref{algo:persons-in-black-areas} gives a different classification (partition) of all individuals stored in the system: $suspect_{t,0}$, $suspect_{t,1}$, ..., $suspect_{t,k}$ (\texttt{SUSPECT}$_t$ set). The unclassified persons are considered uninfected (they have not visited any infectious areas $ba_k \in BA$). Accordingly, the investigation system can be queried about black areas and the people who frequented them.



\section{System implementation}
\label{sec:systemImplemention}

This section describes the small-scale version of our proposed system that we have implemented. The main purpose of this demonstration is to show the applicability of the ideas discussed earlier in the paper. In brief, the implemented system has three essential parts (Figure~10): (1)~mobile phone application for end-users, (2)~automatic investigation mechanism (described in sections~\ref{subsec:data-storage} and~\ref{subsec:data-leveraging}), and (3)~an interface dedicated to the government and health authorities.

\begin{figure}[!b]
\begin{center}
\includegraphics[width=5in]{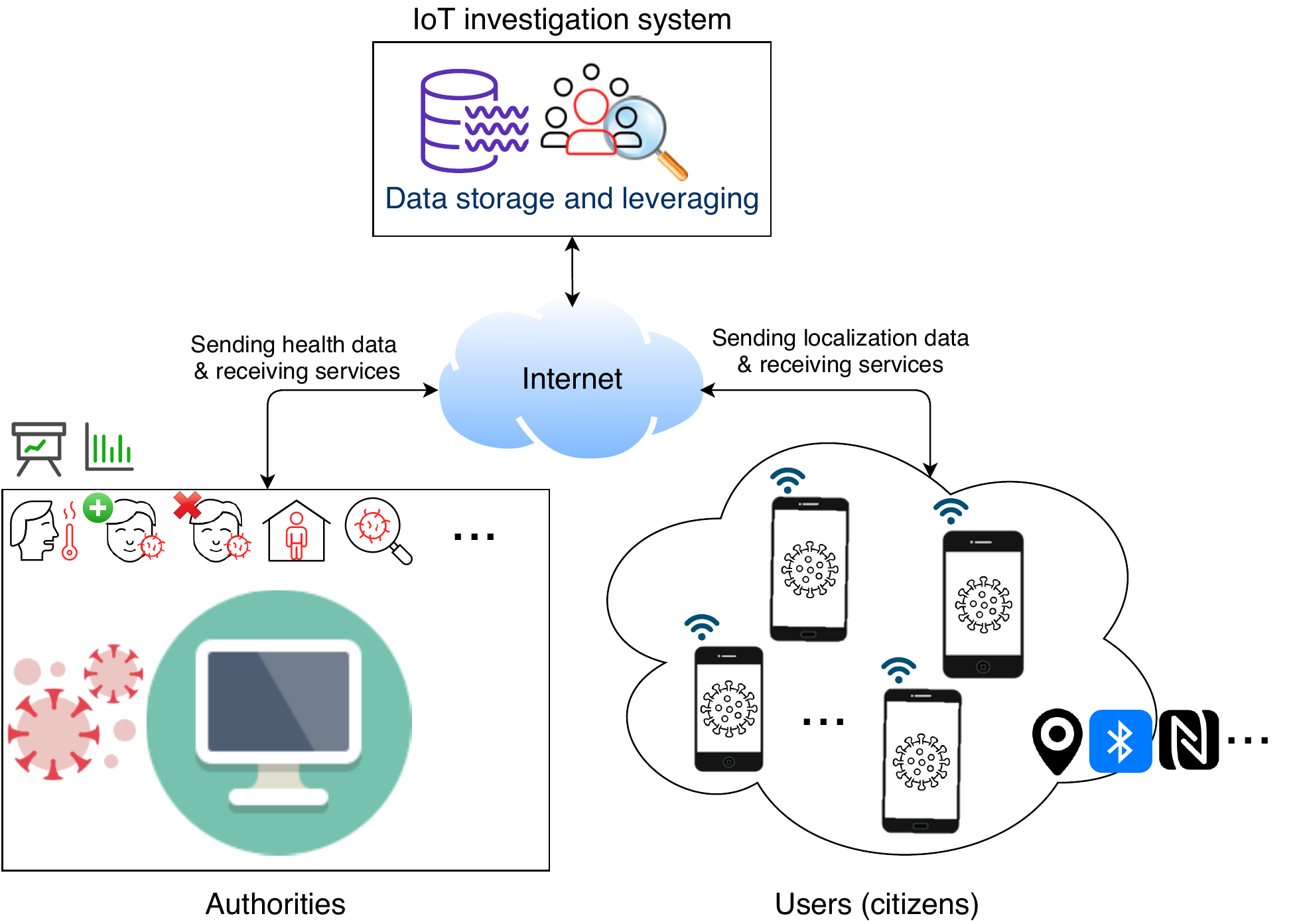}
\caption{Developed application architecture.}
\end{center}
\label{fig:application-architecture}
\end{figure}

\subsection{User application}
\label{subsec:user-application}

The developed mobile application collects the required geo-localization information by sending the GPS longitude and latitude coordinates along with the corresponding time to the investigation system. This operation is repeated periodically every $x$ unit of time. To avoid redundancies (useless data), during this defined period (i.e., $x$ time unit), when a user remains in the same location, his coordinates will not be reported to the system nor stored in the phone local database. Similarly, if a user moves for a certain distance of $y$ meters or less, he will be considered as stationary and his coordinates will not be collected. Also, as previously mentioned, in the event of Internet disconnection, each smartphone will store locally the gathered data and wait for reconnection.

\begin{figure}[!t]
\begin{minipage} {.35\textwidth}
\begin{center}
\includegraphics[width=3in, width=2in]{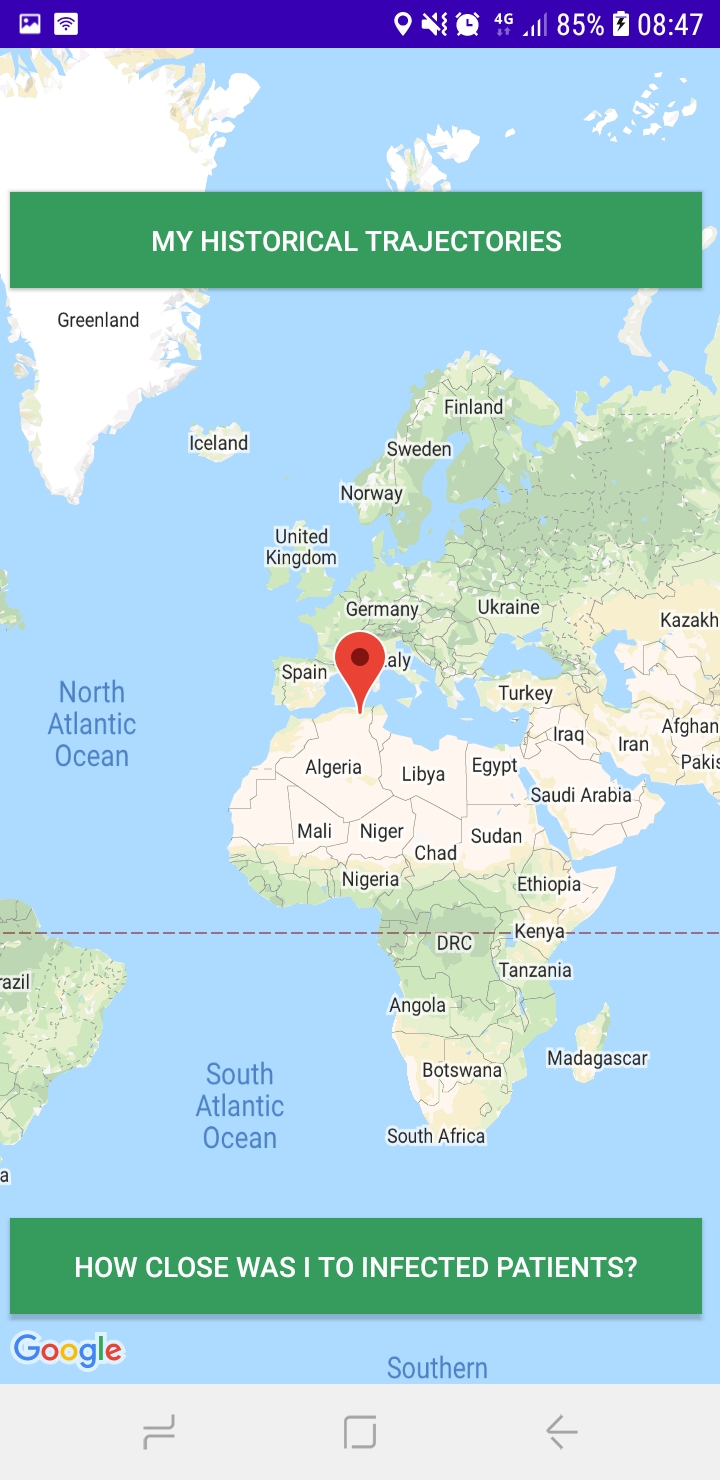}
\end{center}
\end{minipage}
\begin{minipage} {.35\textwidth}
\begin{center}
\includegraphics[width=3in, width=2in]{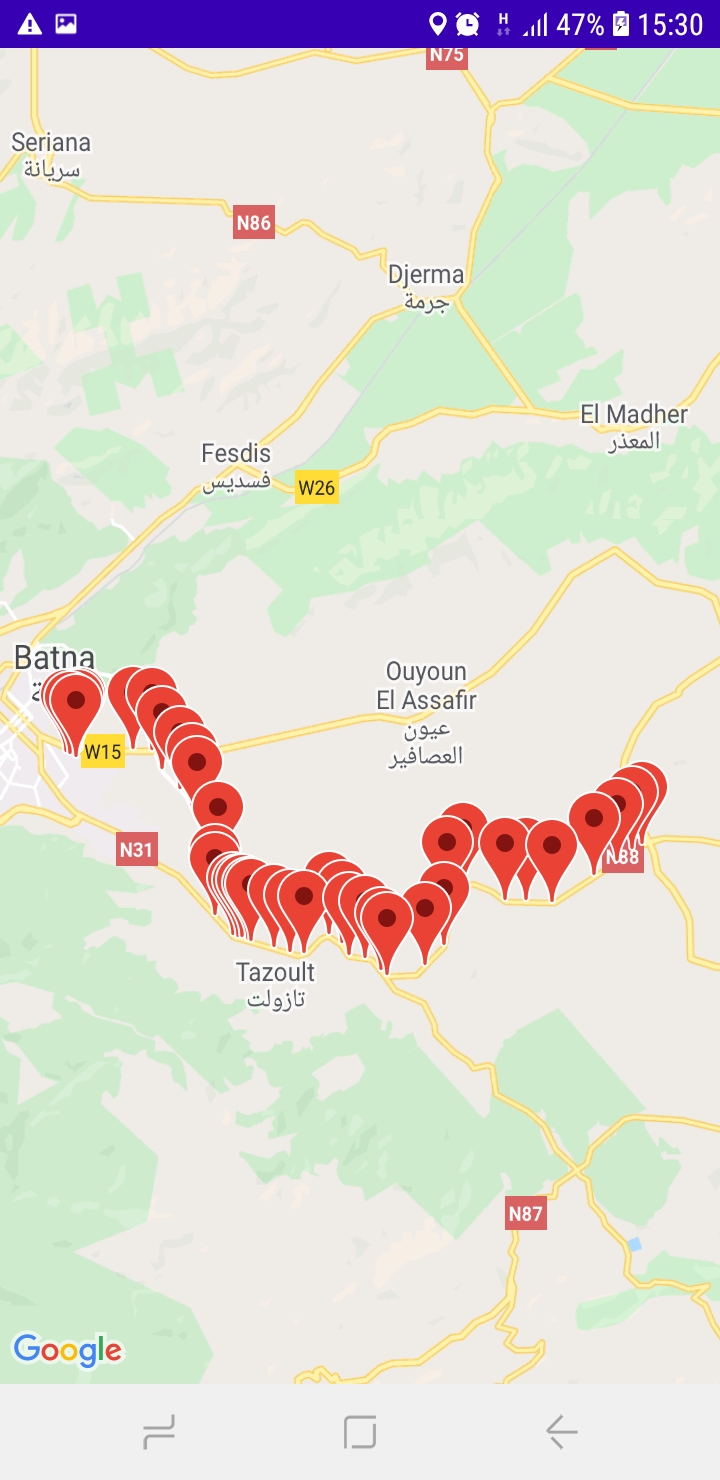}
\end{center}
\end{minipage}
\begin{minipage} {.\textwidth}
\begin{center}
\includegraphics[width=3in, width=2in]{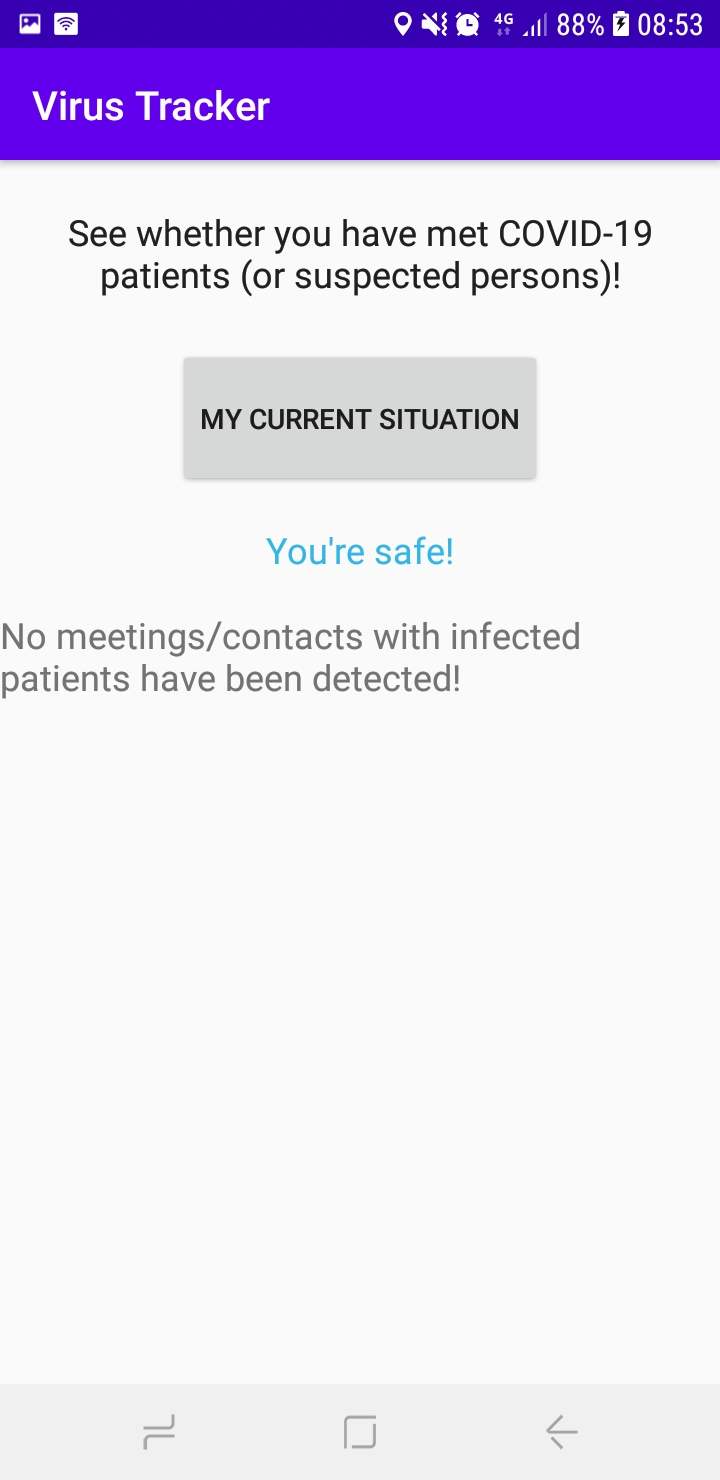}
\end{center}
\end{minipage}
\caption{Screenshots from the developed application.}
\label{fig:app-screenshots}
\end{figure}

The application can offer numerous services to the end-users. For instance, users can review their historical trajectories during any possible correct period that they can define (Figure~\ref{fig:app-screenshots}). A second most important service that can also be offered is allowing end-users to know if they have met infected people.

\subsection{Investigation system}
\label{subsec:globalDatabase}

The investigation system is continuously fed with both the locations of users and health data (e.g., confirmed patients, closed cases, etc.). Figure~12 provides an example of historical trajectories taken by a single user of the system. As previously stated in the paper, the collected locations of persons are stored as immutable append-only data (since there is no need to change the gathered tracks, no updates are allowed). 

Due to the continuous periodic data gathering, an important storing space is needed. However, for demonstration, in this small-scale evaluation version, a single PC was considered to store both health and localization data.

\begin{figure}[!t]
\label{table-trajectories}
\begin{center}
\includegraphics[width=2.5in]{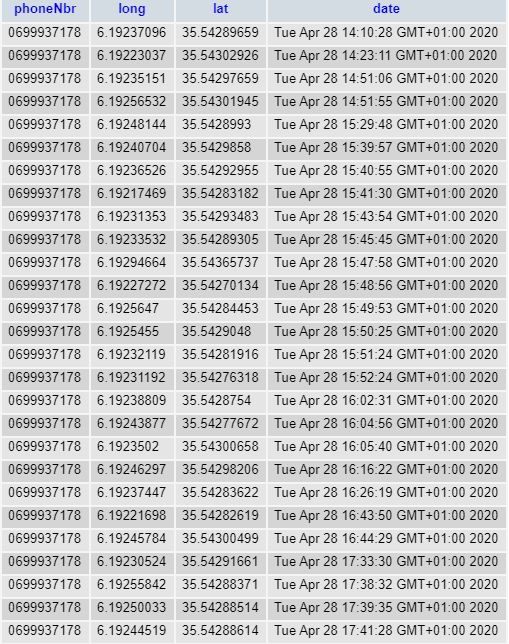}
\caption{Snippet of the collected trajectories of one user.}
\end{center}
\label{fig:Snippet-collected-trajectories}
\end{figure}

\subsection{Authorities interface}
\label{subsec:authorities-interface}

\begin{figure}[!t]
\begin{center}
\includegraphics[width=5in]{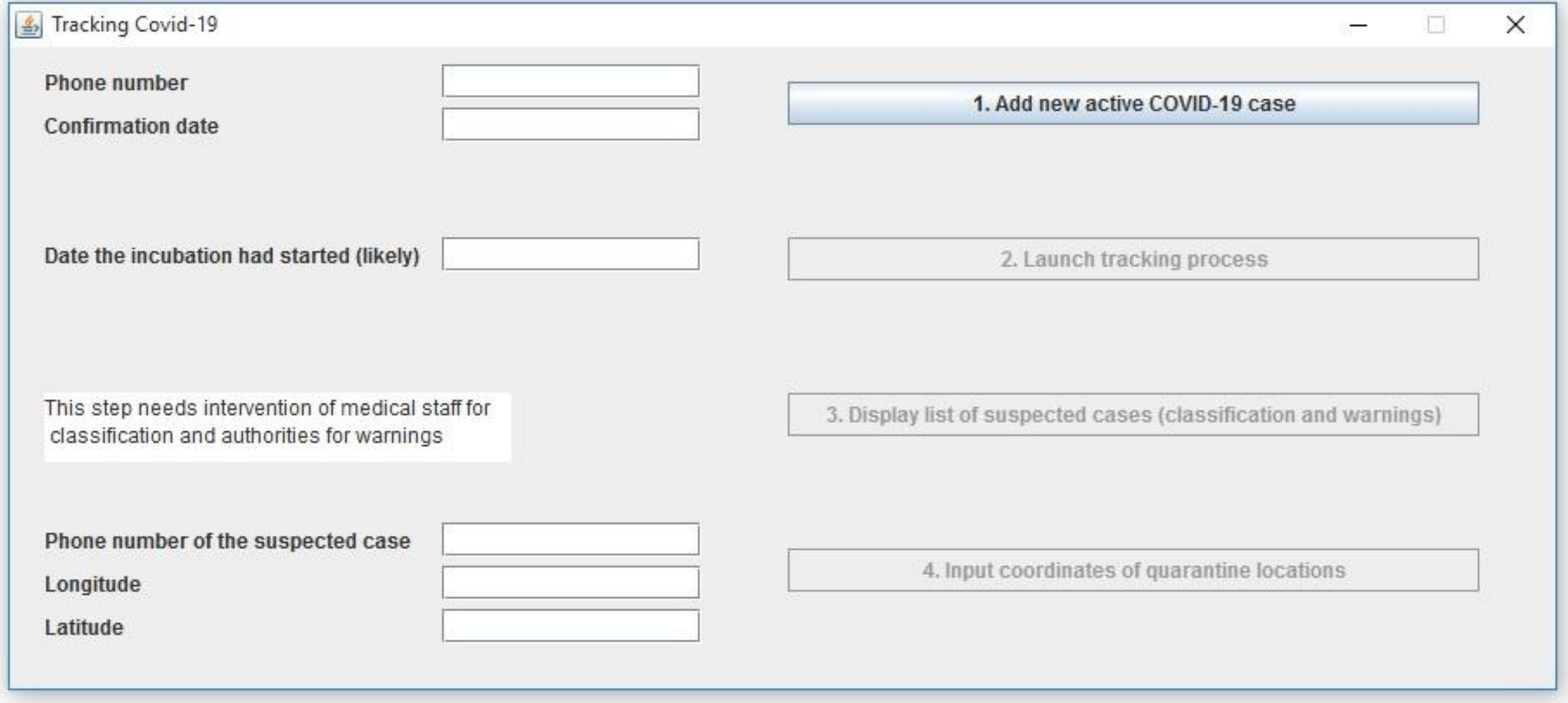}
\includegraphics[width=5in]{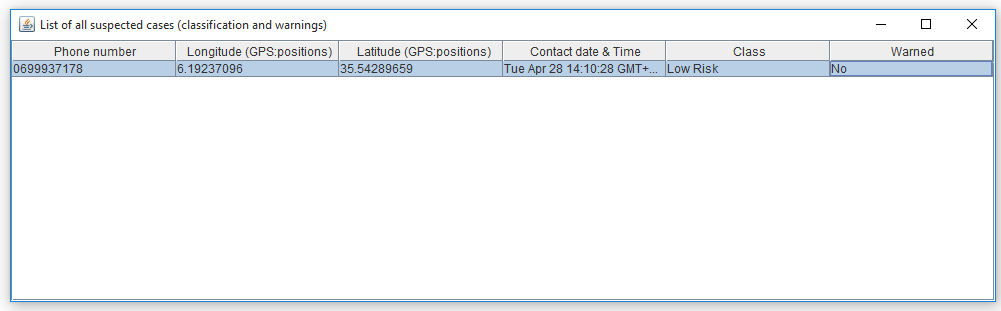}
\caption{Searching for all suspected cases after the report of a new positive case.}
\end{center}
\label{fig:authoritiesInterface}
\end{figure}

Numerous functionalities can be offered to the government, health authorities, and other entities that are responsible for monitoring, evaluating, and controlling the outbreak. The most important functionality is the ability to search for novel suspected cases. To do so, the authorities must first provide the investigation system with the necessary new coronavirus cases, deaths, and recovered patients. Upon discovering and entering a new infected case, the proposed system can find all people who had contact with this patient (Figure~13). 

As previously explained, the proposed system classifies the newfound suspected people into disjoint subsets. For example, if we assume that the newly discovered patient $P_i$ belongs to subset $S_0$, then, all persons who had direct contact with him will belong to $S_{1}$. And, all persons who had direct contact with persons from $S_{1}$ will belong to $S_{2}$, and so forth. This way, categories of high and low risk can be easily determined. Once determined, all suspected cases will be (1)~stored in the system and (2)~informed to immediately start self-isolation at home. The authorities can also apply several techniques to determine if suspected persons are violating the imposed quarantines.  

The investigation system must be designed so that (1)~it helps the authorities in their endeavor of fighting coronavirus, and (2)~it protects the privacy of users. The latter goal can be achieved through the designed GUI and offered functionalities, which must be defined based on a well-established privacy model.

\section{On the mathematical modeling and prediction of infectious diseases}
\label{sec:mathematical-modeling}

Mathematical modeling of infectious diseases is a powerful tool that allows analyzing, understanding, and predicting the behavior of pandemics~\cite{Het00,KM27,DL01}. For instance, these models can be utilized to help authorities set up the best strategies for successful pandemics control. In the following three subsections, we will briefly show how different mathematical prediction models can benefit from the proposed investigation system. More exactly, how can the two generated lists/sets (\texttt{SUSPECTED}$_t$ and \texttt{SUSPECT}$_t$) of (1)~suspected/infected persons and (2)~foci of infection (black areas) help attain a more accurate realistic estimate of disease spreading rate.










\subsection{$\theta$-SEIHRD model}
\label{subsec:theta-SEIHRD-model}

More recently, a new mathematical model, called $\theta$-SEIHRD\footnote{$\theta$-SEIHRD: Susceptible Exposed Infectious Hospitalized Recovered Dead. $\theta$ represents the fraction of infected people that has been detected.}, has been specifically proposed for the coronavirus disease~\cite{IFVR20}. This model assumes that the pandemic spatial distribution inside a territory is omitted. $\theta$-SEIHRD also assumes that people in a territory are characterized to be in one of the following nine compartments: S~(Susceptible), E~(Exposed), I~( Infectious), Iu~(Infectious but undetected), HR~(Hospitalized or in quarantine at home), HD~(Hospitalized that will die), Rd~(Recovered after being previously detected as infectious), Ru~(Recovered after being previously infectious but undetected) or finally D~(Dead by COVID-19). The most important parameters of this model that determine the pandemic spreading are: $\beta^{(i)}_{E}$, $\beta^{(i)}_{I}$, $\beta^{(i)}_{Iu}$, $\beta^{(i)}_{HR}$, and $\beta^{(i)}_{HD}$. These parameters represent the disease contact rates ($day^{-1}$) of a person in both the corresponding compartments ($E$, $I$, $Iu$, ...) and territory ($i$) (without taking into account the control measurements). Indeed, it is very hard to determine these contacts rate without having an idea about who contacted whom. Also, the contacts between persons are not sufficient to estimate the coronavirus contact rates. This disease can be transmitted without direct contact with patients (e.g., touching an object previously used by an infected person). The determination of both sets \texttt{SUSPECTED}$_t$ and \texttt{SUSPECT}$_t$ helps to accurately estimate the contacts rate and to determine the set of \textit{Infectious but undetected} (Iu) persons. In addition to this, our proposed investigation system can also help to determine the pandemic spatial distribution inside any monitored territory. As it is well-known, it is very important to understand diseases spreading, determine the safe and risky areas, and take control measurements with insight.




\subsection{SIR model}
\label{subsec:SIR-model}

SIR\footnote{SIR: Susceptible Infective Removed model.} is one of the most studied mathematical models for the spreading of infectious diseases~\cite{KM27,HK13,DL01}. In this model, the most important parameters that determine the pandemic spreading are $\alpha$, $\gamma$, and $\beta$. Where $\alpha$ is the probability of becoming infected, $\gamma$ is the number of infected people (new infections are the result of contact between infectives and susceptibles), and $\beta$ is the average number of transmissions from an infected person (determined by the chance of contact and the probability of disease transmission). The major difficulty or drawback of this model is estimating these parameters. The answer to this question is our proposed investigation system. 



\subsection{SEIR model}
\label{subsec:SEIR-model}

SEIR\footnote{SEIR: Susceptible Exposed Infectious Recovered model.} is also one of the most important mathematical models for the spreading of infectious diseases~\cite{AS84}. In this model, the main parameter that determines the pandemic spreading is the contact rate $\beta(t)$, which is the average number of susceptibles in a given population contacted per infective per unit time. The main difference between SEIR and SIR lies in the addition of a latency period. More details can be found in~\cite{AS84,BPP14,PKM15,FK14}. Endowing the SEIR model with accurate contact information can help it obtain better results about the spreading (prediction) of infectious diseases. As previously highlighted, this is exactly where our proposed solution comes in handy. It can help to understand and estimate important parameters for numerous mathematical models proposed for infectious diseases.    



\section{System advantages and shortcomings}
\label{sec:advantages-disadvantages}

In this section, we provide the main advantages and disadvantages of the proposed investigation system. First, among the numerous interesting features of this IoT system we mainly cite:

\begin{itemize}

\item This system allows tracking the trajectories of infected persons, and this, days before the appearance of their symptoms. The system also allows determining all persons who were in close contact with infected patients. Thus, the former can be immediately confined and the proper measurements can be rapidly taken by health authorities. Since all persons who have met infected patients can be identified earlier (i.e., parents, friends, colleagues, and most importantly, those who cannot be easily identified using conventional techniques), the proposed tracking system allows an early outbreak control.  
	
\item The proposed system can be utilized to identify black zones, which can be the main source of virus-spreading. As previously explained, a given location is said to be a source of contamination if numerous infected persons have visited it (the intersection zones of high infections can be identified based on the trajectories of infected persons).

\item According to~\cite{ref1}, the coronavirus can live for several hours without a host. To remedy this, the proposed system can be configured to find all uninfected persons who have visited locations that were previously visited by other infected persons (and this, hours after the patients have left this location).
		
\item The proposed system can help reduce the economic damages generated by the suspension of all activities. Instead of shutting down all sports, social and economic activities, the authorities can impose confinement on only a few people. 

\item In addition to coronavirus, the proposed solution can be utilized for any potential outbreak which might be more powerful and have a higher spreading pace.  

\end{itemize}

Despite its advantages, the proposed solution suffers from some issues that must be adequately tackled:

\begin{itemize}

\item For instance, if certain persons do not respect the safety instructions (e.g., they did not intentionally or unintentionally save their trajectories when leaving their houses or cars), it will be impossible to determine whether they have met infected persons. The absence of one or several trajectories does not mean the total failure of the proposed investigation system. Indeed, the efficiency of this system is proportional to its users. If $x$\% of the population has been involved, then approximately $x$\% of the persons who had close contact with patients will be determined. 

\item The proposed system can be empowered with extra functionalities such as machine (deep) learning abilities. Moreover, the system can be utilized to determine whether citizens are respecting social distancing and also to check whether the persons who are suspected to be infected are respecting the imposed confinement. 

\end{itemize}

\section{Conclusions and recommendations}
\label{sec:conclusion}

In this work, we proposed a system that can quickly determine all persons who are suspected to be infected by the coronavirus. While being specific to this virus, the proposed solution can also be applied to control other pandemics. The objective (as specified in the paper) is not tracking people but tracking extremely dangerous viruses. To ensure a practical and more efficient application of this proposal, the following points must be taken into account:

\begin{itemize}

\item \textit{System utilization}: to keep the public health situation under control, it is recommended to launch such an automatic investigation process as early as dangerous rapid infections start taking place. In the case where the outbreak achieves large scales, the operation of detecting, tracking, and surveilling a huge number of infected people might become useless.  

\item \textit{System efficiency}: for a more efficient system, we recommend exploiting all possible existing resources that can enhance the quality of the collected data (cellphones, security camera footage, credit card records, etc.).

\item \textit{Data privacy}: finally, to ensure the privacy and security of the collected data,  which is personal and highly sensitive, an authority of trust must manage this whole process. 
	
\end{itemize}

\bibliographystyle{abbrvurl}
\bibliography{mybibfile}


\end{document}